\documentclass[a4paper,10pt,twocolumn]{revtex4}
\usepackage[english]{babel}
\usepackage{indentfirst}
\usepackage{graphicx}
\usepackage{amsmath}
\usepackage{amssymb}
\usepackage{amsthm}
\usepackage{mathrsfs}
\usepackage{lscape}
\usepackage{bm}
\usepackage{float}
\usepackage{longtable}
\usepackage{graphics}
\usepackage{color}

\begin{document}
\title{Cavity-QED based on collective magnetic dipole coupling: spin ensembles as hybrid two-level systems}
\author{Atac Imamo\~glu}
\affiliation{Institute of Quantum Electronics, ETH Zurich, 8093
Zurich, Switzerland}

\vspace{-3.5cm}

\date{\today}
\begin{abstract}
We analyze the magnetic dipole coupling of an ensemble of spins to a superconducting microwave
stripline structure, incorporating a Josephson junction based transmon qubit. We show that this
system is described by an embedded Jaynes-Cummings model: in the strong coupling regime, collective
spin-wave excitations of the ensemble of electrons pick up the nonlinearity of the cavity mode, such
that the two lowest eigenstates of the coupled spin-wave + microwave-cavity + Josephson-junction
system define a hybrid two-level system. The proposal described here enables the use of spin
ensembles as qubits which can be coherently manipulated and coupled using the same nonlinear-cavity.
Possibility of strong-coupling cavity-QED with magnetic-dipole transitions opens up the possibility
of extending previously proposed quantum information processing protocols to spins in silicon or
graphene, without the need for single-electron confinement.
\end{abstract}
\maketitle

As compared to the coupling of a single emitter,  the strength of optical excitations out of an
ensemble of two-level emitters is enhanced by the square-root of the number of emitters
($\sqrt{N_s}$). This collective enhancement of light-matter coupling in free-space has played a
central role in quantum memory and repeater protocols \cite{dlcz01}. In the context of cavity quantum
electrodynamics (QED), the $\sqrt{N_s}$ enhancement comes at the expense of the desirable
nonlinearity of the coupled cavity-emitter system \cite{lukin99}. Nevertheless, strong
electric-dipole coupling of a number of diverse systems including inter-band excitons
\cite{weisbuch92}, intersubband plasmons \cite{sirtori} and cold-atomic ensembles \cite{esslinger} to
high quality ($Q$) factor optical cavities have been demonstrated: the signature of strong coupling
for these systems is the appearance of Vacuum Rabi-splitting of two dressed modes, each with a
perfectly harmonic spectrum. Direct magnetic-dipole coupling of spins to cavity modes on the other
hand, have been totally ignored, even though collective excitations out of an ensemble of $\sim 10^6$
spins could easily reach a corresponding linear strong coupling regime using the superconducting
microstrip (SCM) cavities recently realized in the context of circuit-QED experiments
\cite{blais04,wallraff04}.

In this Letter, we describe how two-level  hybrid-spin qubits can be defined using  magnetic-dipole
coupling of an ensemble of spins to SCM cavities with a built-in nonlinear element such as a transmon
qubit \cite{schoelkopf07}. We consider a geometry where the transmon qubit, with ground and excited
states denoted by $|a\rangle$ and $|b\rangle$, is introduced at an electric-field maximum of the SCM
cavity. In contrast, an ensemble of spins are placed at a location where the magnetic field is
maximum (Fig.~1). The Hamiltonian of the combined system in the interaction picture is given by
\begin{equation}
\hat{H} = \hbar g_c (\hat{\sigma}_{ba} \hat{a}_c e^{-i\delta t} + h.c.) + \hbar g_m \sum_{i=1}^{N_s}
(\hat{\sigma}_{-}^i \hat{a}_c e^{-i\Delta t} + h.c.) \label{eq1}
\end{equation}
where $\hat{\sigma}_{ba} = |b \rangle \langle a|$ and $\hat{\sigma}_{-}^i$ denotes the spin lowering
operator of the $i^{th}$ spin. $\hat{a}_c$ is the annihilation operator of the SCM cavity-mode. $g_c$
($g_m$) denotes the electric (magnetic) dipole coupling strength of the transmon-qubit (single-spin)
to the SCM cavity mode. We assume here that the transmon-qubit (spins) is (are) red detuned from the
cavity-mode by $\delta$ ($\Delta$).

\begin{figure}[ht]
\includegraphics[width=0.50\textwidth]{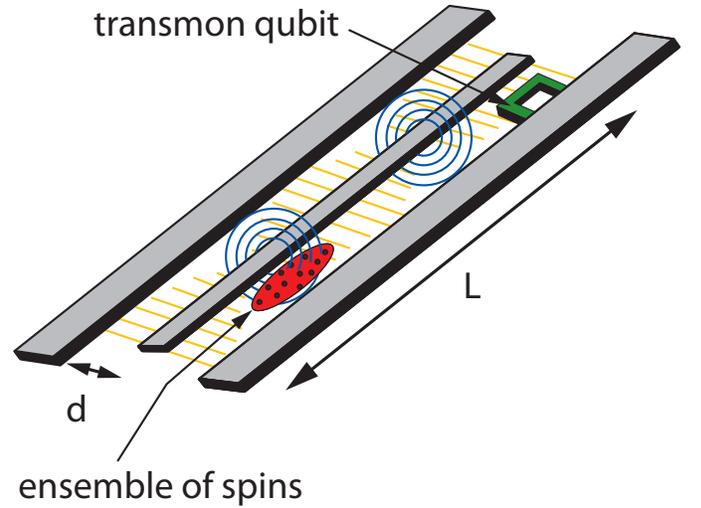}
\caption{\small{Schematic of the system consisting of an ensemble of spins coupled to the quantized
magnetic field of a superconducting microstripline cavity with $d \sim 10 \mu$m and $L \sim 1$cm. The
length of the straight orange lines indicate the strength of the cavity mode electric field. The blue
lines that encircle the center conductor depict the magnetic field lines at the locations where their
strength is maximum. The presence of a transmon qubit at an electric field maximum ensures that the
cavity has a large nonlinearity. The ensemble of spins could either be that of electrons in the
silicon substrate or cold ground-state atoms trapped $\sim 10 \mu$m above the cavity structure.}}
\end{figure}

Before proceeding, we highlight the  recent work discussing a very similar scenario where an ensemble
of polar molecules  are coupled to a SCM cavity containing a josephson-junction qubit
\cite{zoller06,molmer08}. The underlying idea for this pioneering work was the use of strong
electric-dipole coupling strength of polar molecules to achieve an interface between a solid-state
and an ensemble molecular qubit, with the goal of using the molecular qubit as quantum memory. The
present proposal in contrast is based on strong coupling regime of magnetic dipole interaction in
systems where electric-dipole coupling is vanishingly small: such systems have the advantage of being
immune to charge noise and could constitute qubits with much longer coherence times.

A key element of the proposal we describe here is the unprecedented values of cavity-qubit
electric-dipole coupling strength ($g_c$) achieved using a transmon qubits in SCM cavities,
saturating the fundamental limit $g_c = \sqrt{\alpha} \omega_c$ \cite{schoelkopf07}, where
$\sqrt{\alpha}$ and $\omega_c$ denote the fine-structure constant and the bare-cavity resonance
frequency, respectively. In this regime, the condition $g_c \gg g_m \sqrt{N_s}$ is satisfied and we
can start our analysis by neglecting the second term in Eq.~\ref{eq1}: we then obtain the celebrated
Jaynes-Cummings (JC) spectrum for the coupled cavity-transmon system, with an anharmonic spectrum. In
the first part of our analysis, we will focus on the resonant case with $\delta = 0$.

We now turn our attention to the coupling of the {\sl cavity-transmon molecule} to the ensemble of
spins. If we choose $\Delta$ such that $|g_c - \Delta| \ll g_c$ and the cavity Q is high enough to
satisfy $g_c \gg \kappa_c = \omega_c/Q$, we would only need to consider the coupling of collective
excitations of the spins to the two lowest-energy eigenstates of the JC ladder; namely the ground
state $|\tilde{0}\rangle = |a,0_c\rangle$ and the symmetric state of the single-excitation ($n=1$)
manifold $|\tilde{1}\rangle = (|a,1_c\rangle + |b,0_c\rangle)/\sqrt{2}$. This is a consequence of the
fact that the energy separation between the lowest energy states of the one ($n=1$) and two ($n=2$)
photon excitation manifolds in the JC-ladder is given by $g_c (\sqrt{2}-1)$, which is, by our initial
assumption, much larger than the energy scale ($g_m \sqrt{N_s}$) associated with coupling to spins.
Since the eigenstates of the spin ensemble form a harmonic ladder, the coupling of the two-level
system spanned by the eigenstates $|\tilde{0}\rangle$ and $|\tilde{1}\rangle$ to collective spin
excitations maps on to an embedded JC ladder. The scenario we envision is depicted in Figure~2 where
the encircled states form the embedded JC model. Provided that the corresponding coupling strength,
given by $g_m \sqrt{N_s}$, is larger than the decoherence rate of the spin excitation
($\gamma_{spin}$), transmon qubit ($\gamma_{JJ}$) and the SCM cavity-mode ($\kappa_c$), the two
lowest energy states of the embedded JC ladder define a two-level hybrid spin qubit. For $ \delta =0$
and $\Delta = g_c$, the Hilbert space of this qubit is spanned by the states
\begin{eqnarray}
|\tilde{\tilde{0}}\rangle &=& |{\cal G}, a,0_c\rangle  \nonumber \\
|\tilde{\tilde{1}}\rangle &=& \frac{1}{\sqrt{2}}|{\cal E}, a, 0_c\rangle + \frac{1}{2}|{\cal G}, b,
0_c\rangle + \frac{1}{2}|{\cal G}, a, 1_c\rangle
 \label{eq2}
\end{eqnarray}
where $|\cal{G} \rangle$ and $|\cal{E}\rangle$ = $(\sum_{i=1}^{N_s} \hat{\sigma}_{-}^i)$$|\cal{G}
\rangle$$ /\sqrt{N_s}$ denote the fully polarized ground and first excited (single collective
spin-flip) states of the spin ensemble. To estimate the validity of the two-level approximation, we
compare the degree of the anharmonicity given by $g_m \sqrt{N_s} (\sqrt{2}-1)$ with the decoherence
rates typical for the system: $\kappa_c \simeq 1 \times 10^6 \simeq \gamma_{JJ} \ge \gamma_{spin}$.
The single-spin magnetic-dipole coupling strength is given by $g_m = \mu_B \sqrt{\mu_o
(\omega_c-g_c)}/\sqrt {2 \hbar V_c}$, where $\mu_B$ is the Bohr magneton, $\mu_o$ is the vacuum
permeability and $V_c$ is the SCM cavity-mode volume. For a cavity-volume $V_c = 10^{-12} m^2$
(corresponding to a separation of $\sim 10 \mu$m between the center conductor and the ground planes),
$\omega_c = 2\pi \times 10 GHz$, we find $g_m = 1 \times 10^3$ rad/sec. This estimation implies that
we would need $N_s \ge 10^6$ for the two-level approximation to be valid. A $10 \mu$m thick
bulk-doped silicon with a dopant density $n = 10^{16} cm^{-3}$ and an area of $10 \mu$m by $100\mu$m
would yield $N_s \sim 10^8$ spins.

\begin{figure}[ht]
\includegraphics[width=0.5\textwidth]{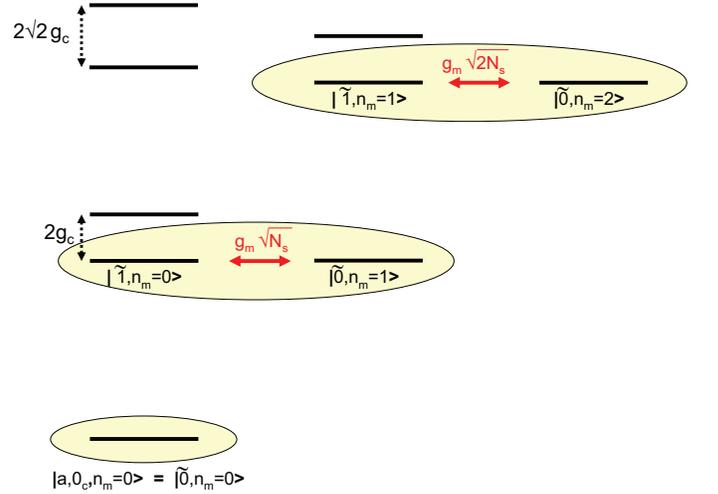}
\caption{\small{The level diagram depicting the lowest energy eigenstates of the
dressed cavity-transmon system along with the lowest energy excited states of the spin ensemble. The
anharmonicity of the cavity-transmon molecule is large enough to ensure that only a subset of
eigenstates, highlighted in yellow, couple non-perturbatively via the collectively enhanced magnetic
dipole transitions. The diagonalization of magnetic dipole interactions among these states yields an
embedded Jaynes-Cummings ladder.}}
\end{figure}

The hybrid qubit defined by the computational states $|\cal{G} \rangle$ and $|\cal{E}\rangle$ brings
up a number of benefits, such as collective enhancement of coupling to external coherent fields, the
possibility of working with spin systems with weak spin-orbit interaction and avoiding the
requirement to isolate a single spin. However, using the anharmonic spin-wave excitations as a qubit
would require that the ensemble of spins are either completely polarized or are in a mixture of
dark-state that could not be further polarized due to the symmetry
\cite{imamoglu03,taylor03,taylor05}. Fortunately, the same strong cavity coupling utilized in
defining the hybrid qubit could also be used to cool down the spin ensemble into a mixture of dark
states. This would be achieved ideally in the limit where the cavity-Q is reduced and the spin
excitation is transferred to phonons via non-radiative cavity dissipation. Provided that the
cavity/substrate temperature satisfies $T_c \le 70$mK, it should be possible to ensure that the spins
are in a mixture of dark states with mean number of collective excitations much smaller than unity.
However, the inhomogeneous broadening due to different collective coupling strengths associated with
different dark states may become a limitation if the degree of spin polarization is not high
\cite{taylor05}. A temporary reduction in cavity-Q could be achieved by laser irradiation that
induces ohmic losses. We emphasize that the initial temperature of the spins need not be low.

Coupling of spins to the cavity mode  would require that the Zeeman splitting $\omega_z$ satisfies
$\omega_z = \omega_c - g_c$, which in turn implies the presence of an effective magnetic field of $B
\sim 0.5$Tesla. Such magnetic fields will substantially reduce the cavity-Q. Ideally, this problem
can be remedied by inducing the electron spin-splitting via polarized nuclear spins of the host
material: it is well known that dynamical polarization techniques lead to effective Overhauser fields
that can exceed 2 Tesla and that survive for more than a minute after active polarization is turned
off. Since these effective fields are predominantly due to Fermi-contact hyperfine interaction,
cavity-Q would remain unaffected.

Alternatively, we could envision a local external field of $B \sim 1$Tesla that vanishes at the site
of the transmon qubit. Such an external field will reduce the cavity-Q and the condition $g_c \gg g_m
\sqrt{N_s} > \kappa_c$ may no longer be satisfied. However, provided that $Q > 100$, $g_c \gg
\kappa_c$ is still satisfied and it is possible to overcome the limitation of a lossy cavity: to this
end, we assume that the transmon qubit is detuned from the cavity mode in a way to ensure that it is
resonant with the electron spins ($\Delta = \delta + g_c^2/\delta$). In the limit $\hbar |\delta|
\gg$ all other energy scales, we use a Schrieffer-Wolff transformation to eliminate the first term in
Eq.~\ref{eq1} and find the transformed Hamiltonian to lowest order
\begin{equation}
\hat{H} = \hbar \frac{g_m g_c}{\Delta} \sum_{i=1}^{N_s} (\hat{\sigma}_{-}^i \hat{\sigma}_{ab}  +
h.c.) \label{eq3}
\end{equation}
where we assumed that $\delta$ is large enough to ensure that $|\delta - \Delta| \ll |\Delta|$ and
that $g_m \sqrt{N_s} \ll g_c$. The transformed Hamiltonian is also of the JC-form since the
collective spin lowering operator $(\sum_{i=1}^{N_s} \hat{\sigma}_{-}^i)$$ /\sqrt{N_s}$ approximates
a bosonic creation operator in the limit $N_s \gg 1$ for low excitation manifolds. The condition for
strong coupling in this case is given by $g_c g_m \sqrt{N_s}/\Delta
> \kappa_c g_c^2/\delta^2, \gamma_{spin}, \gamma_{JJ}$: in this limit, the cavity-mode acts as a
quantum-bus and its virtual excitations mediate the long distance JC interaction between the transmon
qubit and the collective spin excitations.

Having demonstrated that the two  lowest energy collective excitations of a spin ensemble could
define an anharmonic two-level system, we address the possibilities for manipulation of quantum
information. Arbitrary unitary rotations of the two-level system defined by the states
$|\tilde{\tilde{0}}\rangle$ and $|\tilde{\tilde{1}}\rangle$ could be effected by an external resonant
ESR field or by Raman transitions induced by two phase-locked laser fields. In order to couple
spin-ensemble qubits, one could consider a cavity containing several transmon qubits, each with a
different detuning. By swapping the spin qubits to and from the same transmon qubit, it is possible
to effect square-root of swap operation \cite{loss98}. Each transmon qubit will then constitute a
different channel for parallel implementation of two-qubit gates \cite{imamoglu99}.

An advantage of  ensemble spin qubits over that of transmon qubits is their potentially much longer
decoherence times. This would particularly be true for ensemble spins in pure silicon-28 with no
hyperfine interaction and ultra-small spin-orbit coupling. In GaAs, the ensemble spin approach will
enable conversion of quantum information carried by the spin/transmon system to that of a propagating
photon. As compared to GaAs single-electron spin qubits, ensemble spins have the advantage of reduced
hyperfine decoherence \cite{loss02}. It should be emphasized however, that spin-orbit induced
decoherence eventually becomes prominent in bulk GaAs structures and limits the spin coherence time
to $\sim 1 \mu$sec \cite{awschalom98}.

Another interesting system for which the formalism discussed here applies is an ensemble of cold
atoms trapped $~10 \mu$m above the SCM cavity structure. At such distances, the adverse effects of
the solid-interface on ground-state atoms can be avoided \cite{esslinger-private}. Without the need
for strong external magnetic fields, the magnetic dipole coupling of the hyperfine transition of the
atomic ensemble to the nonlinear cavity mode will then lead to an anharmonic energy level diagram for
the collective atomic (hyperfine) spin excitations. In addition to providing a long-lived memory, a
cigar shaped atomic ensemble coupled to a cavity-transmon system would enable near-unity efficiency
conversion of a microwave photon to an optical photon that can be collimated using a low numerical
aperture lens \cite{dlcz01}.

In conclusion, we demonstrate that collective enhancement of magnetic dipole coupling of an ensemble
of spins would lead to the strong coupling regime of cavity-QED. The presence of a transmon qubit in
the SCM cavity would lead to the  realization of an embedded JC model where the harmonic emitter
(spin) system picks up the nonlinearity of the cavity. In the context of quantum information
processing, our findings generalize the conclusions previously drawn for polar molecules to a larger
class of systems, ranging from trapped ground-state atomic gases to spins in bulk silicon or
graphene.


This work was supported by QSIT at ETH Zurich. We wish to
acknowledge many useful discussions with H. Tureci, A. Wallraff and
T. Esslinger.

\end{document}